# Ontology *for* Conceptual Modeling: Reality of What Thinging Machines *Talk About*, e.g., Information

Sabah Al-Fedaghi*
*Computer Engineering Department*
*Kuwait University*
*Kuwait*
salfedaghi@yahoo.com, sabah.alfedaghi@ku.edu.kw

*Abstract* – In conceptual modeling (CM) as a subdiscipline of software engineering, current proposed ontologies (categorical analysis of entities) are typically established through whole adoption of philosophical theories (e.g. Bunge's). In this paper, we pursue an interdisciplinary research approach to develop a diagrammatic-based ontological foundation *for* CM using philosophical ontology as a secondary source. It is an endeavor to escape an offshore procurement of ontology from philosophy and implant it in CM. In such an effort, the CM diagrammatic language plays an important role in contrast to dogmatic philosophical languages' obsession with abstract entities. Specifically, this paper is about developing a descriptive (in contrast to formal) ontology that a modeler accepts as a supplementary account of reality when using thinging machines (TMs; i.e. a reality that uncovers the ontology of things that TM modeling discusses or "talks about," akin to the ontology of natural language). Although existence is a well-established notion, we defend subsistence (Stoic term) as a supplementary mode of reality (e.g. reflection of event). The aim here is aligned toward developing CM notions and processes that are firm enough. Classical analysis of being *per se* (e.g. identity, substance, classes, objects) is de-emphasized in this work; nevertheless, philosophical concepts form an acknowledged authority to compare to. As a case study, such a methodology is applied to the notion of information that provides a common-sense understanding of the world. This application would enhance understanding of the TM methodology and clarify some of the issues that shed light on the question of the nature of information as an important concept in software engineering. Information is defined as *about* events; that is, it is about existing things. It is viewed as having a subsisting nature that exists only through being "carried on" by other things. The results seem to indicate a promising approach to define information and understand its nature.

*Index Terms - Conceptual model, ontology, diagrammatic modeling language, existence, subsistence, what is information*

## I. INTRODUCTION

Ontology has been an increasingly critical notion in computer science and software engineering. "Ontology" refers to the study of the kinds and structures of things, events, processes, and relations in reality. Specifically, in software engineering, systems must be grounded in adequate representations of the targeted domain (portion of reality) to meet the requirements and ensure the quality of the system according to the user´s perspective.

---
*Retired June 2021, seconded fall semester 2021/2022

In this context, conceptual modeling (CM) is pivotal in providing understanding and communicating the meaning of the system's entities and processes. For example, it is claimed that the Bunge-Wand-Weber ontology provides a theoretical basis for appraising modeling practices and the capacity of representation languages. As another example, the phenomenological foundational ontology is said to model entities as phenomena representing mind objects, based on the comprehension of existence in a metaphysical sense, as in the philosophical school of phenomenology [1].

### A. Alternative Approach in This Paper

Current proposed ontologies in CM are obtained through whole adaptation of philosophical theories (e.g. Bunge's; see [2-5]). The resultant ontologies have shown a great diversity of philosophical ideas and theories, yet "it is not clear which approach the modeler should choose and what the basic elements of systems are that need to be represented" [3]. In this paper, we pursue an alternative approach to develop a diagrammatic-based ontology *for* CM in software engineering using philosophical ontologies as a secondary source.

The aim is not metaphysical, which requires a commitment to a certain theory of the world, such as object orientation (e.g. Bunge's) or process orientation (e.g. Whitehead's); rather, the aim is to develop a descriptive (in contrast to formal) ontology that a modeler accepts it as a "ground" of the thinging machine (TM) modeling (i.e. things that TM modeling discusses or "talks about") akin to ontology implicit in natural language. For example, from the CM point of view, it is not an issue as to whether the ontology has some meaningful end toward which it is oriented, besides its practical usage. Thus, the goal is a "being"-oriented ontology (of an irreconcilable heterogeneous world) that supports the diagrammatic modeling language. Nevertheless, this does not mean that systematic categorical examination of the diverse manifestation of being in reality was not followed.

Such an approach is not new; for example, UML ontology has been proposed as a shared public view of a domain in terms of a static model consisting of a class diagram to depict the classes in the domain and their relationships and an object diagram to show particular named instances of those classes [6]. UML had been proposed to be used as an ontology representation language and the OMG requested for proposals in this context [7].

In the proposed approach, diagrammatic languages play an important ontological role in contrast to abstract philosophical analysis. The resulting endeavor is to escape offshore procurement of ontology from philosophy and fitting it in CM. For example, within modern philosophical tradition, "reality" has often been defined as "capability of acting" [8], although this paper is concerned with the reality of *event*s and their *regions* (to be defined in section 2). Specifically, this research is about developing an ontological foundation *for* CM in software engineering based on TMs as a tool for conceptual analysis [9-10]. The intent is to introduce a CM nonphilosophical approach; nevertheless, the work seems to be verging on the philosophical because it is not possible to escape philosophy completely, which might serve to recommence deeper CM research. TM is a starting point for what we anticipate to be a long-term endeavor. The aim is not philosophical; rather, it is aligned toward developing reasonably firm CM notions. Accordingly, classical analysis of being *per se* (e.g. identity, substance), is de-emphasized; nevertheless, philosophical concepts form an acknowledged field of authority to compare to.

### B. Information as a Case Study

As a case study, such a method is applied to the notion of information that provides a common-sense understanding of the world. With the multiple definitions for information, the aim in this paper is to examine the notion of information in CM as part of an ongoing research project, called TM modeling, which utilizes diagrammatic language as an instrument in analyzing different notions. The intended benefits of such an examination include the following:
- Enhance understanding of the TM CM that has been pursued over several years of research.
- Clarify some of the issues that shed light on the question of the nature of information as an important concept in computer science and software engineering.

Specifically, the claimed thesis, in this paper, is that information can be isolated ontologically as a *subsisting* (potential) thing that can *exist* only through being "carried" by embedding it into existing things (e.g. signals).

### C. The Subsistence Thesis

The investigation of reality has been subject to drastic changes in the course of history (e.g. Plato, Newton). Here, we briefly describe the TM reality ontologically in order to provide an intuitive ontological picture of the TM model. This view of reality will be defended in sections 3 and 4.

The TM modeling is a two-level subsistence/existence scheme with origin that goes back to the Stoic modes of reality. Fig. 1 defines the categorical structure of TM modeling. It has two kinds of things and two modes of reality. The basic thesis of this structure is as follows:
(1) Existence is the realm of events (to be defined later).
(2) Events are of two types: sensed (material) events and dependent events that emerge from the sensed events.
(3) Events have footprints.
(4) Existence and subsistence form the realm of reality.

This type of ontology is centered on events, as will be expounded on later in this paper. Accordingly, we claim that the *static* TM level represents the ontic (pre-categorical) reality.

As an *initial* illustration, consider the example (see Fig. 2) that involves the modeling process of *traffic* that becomes into *existence* through being "carried on" cars and roads, and so on. In such a model, the *traffic per se* (as a thing distinguished from cars, roads, traffic lights, etc.) is an immaterial thing and is not a mind-dependent or subjective mode of (everyday) reality. Traffic is something over and above its constituents of cars and roads, or it is a whole that is more than the sum of the parts. It is both autonomous and is irreducible to the constituents from which it emerges.

Traffic subsists when all cars are grounded because activities are halted (e.g. COVID-19 pandemic). This subsistence certainly had "causal powers" on our normal activities. In this case, traffic is a *process in subsistence* that becomes actualized again when the "activities halt" is lifted.

Real *existing* things are those thimacs in the temporal state of nature and are capable of causing effects (create, process, etc.) in other things. As mentioned previously, there are two types of these real existing things: (a) sensed or material things (e.g. cars, roads) and (b) things that are "carried on" things in (a), for example, traffic. Both types have two modes of reality: existence and subsistence.

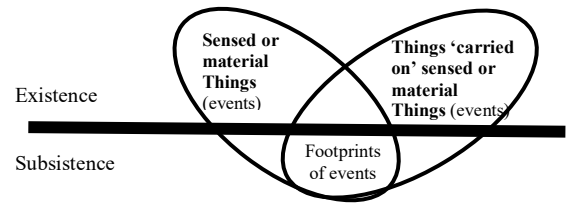

Fig. 1 Two kinds of things and two modes of reality

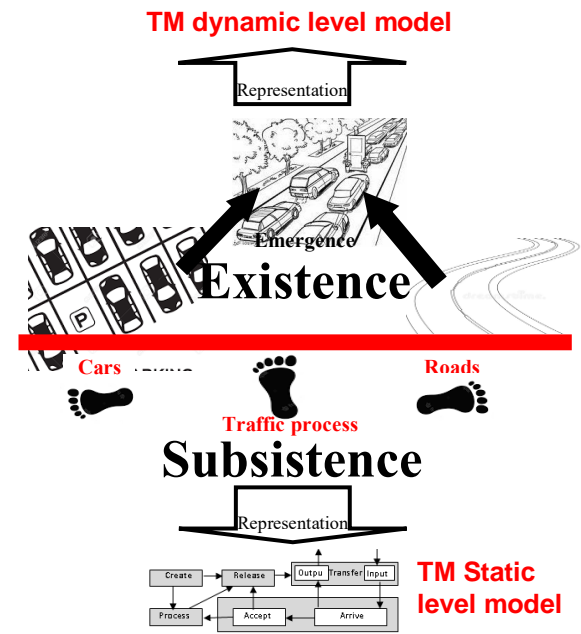

Fig. 2 Subsistence, existence, and TM representation

The subsisting (timeless) traffic turns into the actual (existing) traffic, simply when its corresponding containing events are "spun on"; that is, actual cars are permitted to move. The subsistence of a thing "is due to underlying body, not independent of it […] Should the corporeal cosmos disappear, the extra-cosmic [subsisting thing] would cease to subsist as well" (from discussions of the Stoic ontology in [11]).

Traffic is not itself a solid body, but it is nonetheless *real* when cars and roads (without loss of generality, we ignore other elements; e.g. nonexistent things and mental things) exist; thus, its emergence into existence occurs. Subsistence is a mode of reality on the edge of existence. This emergence is a stronger criterion for existence than a mere potentiality. This is illustrated in Fig. 2. In the figure, the traffic may no longer exist (e.g. COVID-19 pandemic), even though cars and roads exist. However, the traffic still subsists in reality, waiting to emerge in existence. During the COVID-19 pandemic, the traffic is "on hold," subsisting as a *real* phenomenon.

Traffic is in reality because "ingredients" (necessary conditions) of the existence of the traffic are "there" for it to emerge in existence. The traffic, cars, and roads form a nested (e.g. cars emerge from its parts) hierarchy of things at different planes of an ontological organization. Such a hierarchy involves the emergence of existing things (events) from subsisting things (regions), which, in turn, emerges from first occurrences of existing things (events). The sequence of emergencies was interrupted on the border line of emergent things and "bottom" things. Later, we will discuss how the world of multiplicity of subsisting things is created.

### D. Outline of This Paper

Section 2 of this paper describes previous work on TM modeling. The section provides a background for the remaining sections with the aim of a self-contained paper even though certain parts of the section contain new materials that enhance some TM notions. This section also introduces a new example of applying the TM modeling to a case study of a *smart factory* modeled previously in BPMN 2.0.

Sections 3, 4, and 5 comprise studying our own ontological foundation for CM (briefly introduced in the traffic example) using ideas from classical philosophical ontology. The material in the section aims to support the ontological base of our two-level modeling in terms of subsistence and existence modes of reality. Although existence is a well-established notion, the aim of this section is to defend subsistence as a mode of reality. Our thesis is that subsistence is an immaterial (free of matter) configuration (structure constituent and actions) of existing things (i.e. a mirror image [footprints] of events)—for example, traffic emerging from cars, roads, and movements assembled in a certain way (TM static model).

The basis for the thesis in this paper is that the TM static model inscribes things like traffic (and information). Such unobservable things are very common in physics (e.g. strings, extra dimensions, parallel universes).

Sections 6 and 7 apply the theoretical discussion of sections 3-5 to the study of the notion of information with a case study. Information is viewed as a thing *about* events. Information is conceptualized as an immaterial thing (just like traffic) that can be "carried on" existing materials (see Fig. 3).

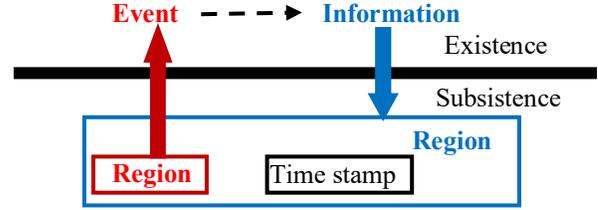

**Fig. 3 Information as an immaterial thing generated as a footprint of an event**

### II. TM MODEL

The TM model is a diagrammatic two-level scheme (dynamic existence and static subsistence) that reflects a targeted portion of reality (i.e. domain within the organization reach). The static level is a mirror image of events in the existence level. In this research work, metaphysics matters concerning the reality of such a diagrammatic scheme as *a whole* (an ontological block) will not be discussed directly, analogous to not involving the issue of reality of mathematics in a mathematical paper, even though the paper may include reality matters (e.g. prices, financing).

In the TM model, "what is there?" is a world of *thimacs* (*thi*ngs/*mac*hines; i.e. a network of thimacs that articulate the furniture of the world). The world is made of thimacs that interconnect with thimacs. **Hereafter, a thimac may be referred to as a thing or machine**. It has a dual nature of being as a *thing* and simultaneously as a *machine*. A thimac is a *machine* when it acts on other thimacs, and it is a *thing* when it is the object of actions by other thimacs. A machine *things* (Heideggerian term); that is, it creates, processes (changes), receives, transfers, and releases. There are two types of creation: (a) initially created (existing) thimacs that are given by the modeler and (b) thimacs that are created by processing other thimacs (e.g. traffic from cars and roads).

### A. The TM Machine

The thimac *machine* executes five actions: *create*, *process*, *release*, *transfer* and *receive* (see Fig. 4). Thimacs are realized through creating, processing, releasing, transferring, and/or receiving thimacs.

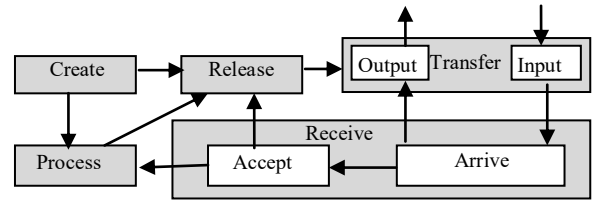

**Fig. 4 Thinging machine**

The thimac *thing* is whatever is created, processed, released, transferred, and received. A thimac is a *machine* that creates, processes, releases, transfers, and receives. TMs' actions are described as follows:

1) *Arrive:* A thing arrives to a machine.
2) *Accept:* A thing enters the machine. For simplification, we assume that arriving things are *accepted* (see Fig. 4); therefore, we can combine the *arrive* and *accept* actions into the *receive* action.
3) *Release:* A thing is ready for transfer outside the machine.
4) *Process:* A thing is changed, handled, and examined, but the thing is not transformed into a new thing.
5) *Transfer:* A thing is input into or output from a machine.
6) *Create:* A new thing is manifested in a machine. In TM, "create" has two senses: (becoming) realized and (being) real. An example of the first sense is a thing that comes into existence as the result of some processing of other things (emergence). In the second sense, a thing is declared as an element of the domain's "inventory."

Additionally, the TM model includes a *triggering* mechanism (denoted by a dashed arrow in this article's figures) that initiates a (nonsequential) flow. Moreover, each action may have its own storage (denoted by a cylinder in the TM diagram). For simplicity, we may omit *create* from some diagrams because the box representing the thimac implies its being-ness (in the model). Additionally, the surrounding box of a machine may be omitted.

### B. Two-level Modeling

The TM involves two *vertical* representations of a thing over a single model. Instead of the common approach of separate diagrammatic representation of static and dynamic features (e.g. class vs. state diagram), a TM language assembles a model that has vertical dynamic representation over static representation. Staticity refers to *timelessness*. The static TM model is built from subsisting *regions* with a logical order imposed by potential flows and triggering. The static model comprises fixed parts, and it simply *subsists.*

Before going into an ontological discussion about existence and subsistent, we present, next, an example of TM modeling to demonstrate the TM diagrammatic representation in a real domain.

### C. Example of TM Modeling

Reference [12] presents an interesting modeling of *IoT-driven* processes. The research problem concerns existing process modeling and process-execution languages (e.g. BPMN 2.0) that are unable to fully meet the IoT characteristics (e.g. asynchronicity, parallelism) of IoT-driven processes. For example, no visual distinction can be made between IoT-related service tasks and non-IoT-related ones. Reference [12] gives a case study that involves a smart factory, as shown in Fig. 5.

Fig. 6 shows the corresponding static TM model developed according to our understanding of the given factory model. In Fig. 6, a container is received by the factory (number 1). This triggers the transmission of information about this arrival to the satellite (2).

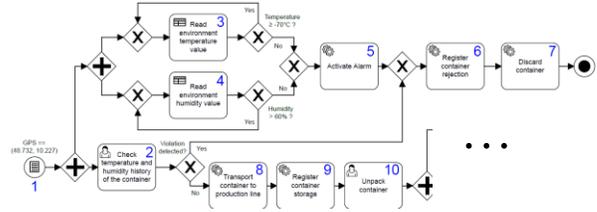

**Fig. 5 IoT-driven business process modeled in terms of BPMN 2.0 (From, partial [12])**

Additionally, the container is processed [3] to extract its history [4]. Note that this extraction involves transfer/receive; that is, the arrival of the container implies the arrival of its history. The history is sent to the supervising worker (5) who processes it (6) in order to decide (according to the detection [or not] of violation) on whether to send the container to the production lane (7) or chemical storage (8) accordingly:

- Releasing the container (7) to the production lane results in registering the container storage (9) and sending it to the unpacking area (10). Then a worker unpacks the container and sends it to a process of sorting and coding (12) in order to move it to the chemical storage (13 and 14).
- Releasing the container to the chemical storage (8 and 15). The container is released (16) after registering its rejection (17) and is transferred (18) to be disregarded.

Inside the chemical storage, the temperature and humidity ((19, 20) and (21, 22), respectively) are monitored continuously. If both failed (23), an alarm is triggered (24), and the container is released (16) after registering its rejection (17) and is transferred (18) to be disregarded.

Development of the dynamic TM model requires the notion of an event. A TM event is defined in terms of a region, a subdiagram of the static model, and time. Thus, events are regions in time, and regions are the interior of events. Events are nothing without their *regions*, and a region cannot exist without its events.

No event is ever identical with another (Davidson), and events are basic particulars in the world (McHenry). For example, Fig. 7 shows the event *The container history is extracted and sent to the supervising worker.* For the sake of simplification, we will represent any TM event by its region. Fig. 8 shows the dynamic TM model where events are represented by their regions. Note that the boundaries of events (maybe the generic five TM action or higher-level events) are selected by the modeler. To save space, the English description of the events is not listed. Fig. 9 shows the chronology of its events.

### III. A GLIMPSE ON NATURE OF THINGS VARIETIES

This section provides a brief introductory discussion on various types of things in reality and their relationship with each other. Specifically, the focus is on subsisting and existing things that form the TM categorical organization in preparation for an ontological discussion of these notions in the next section.

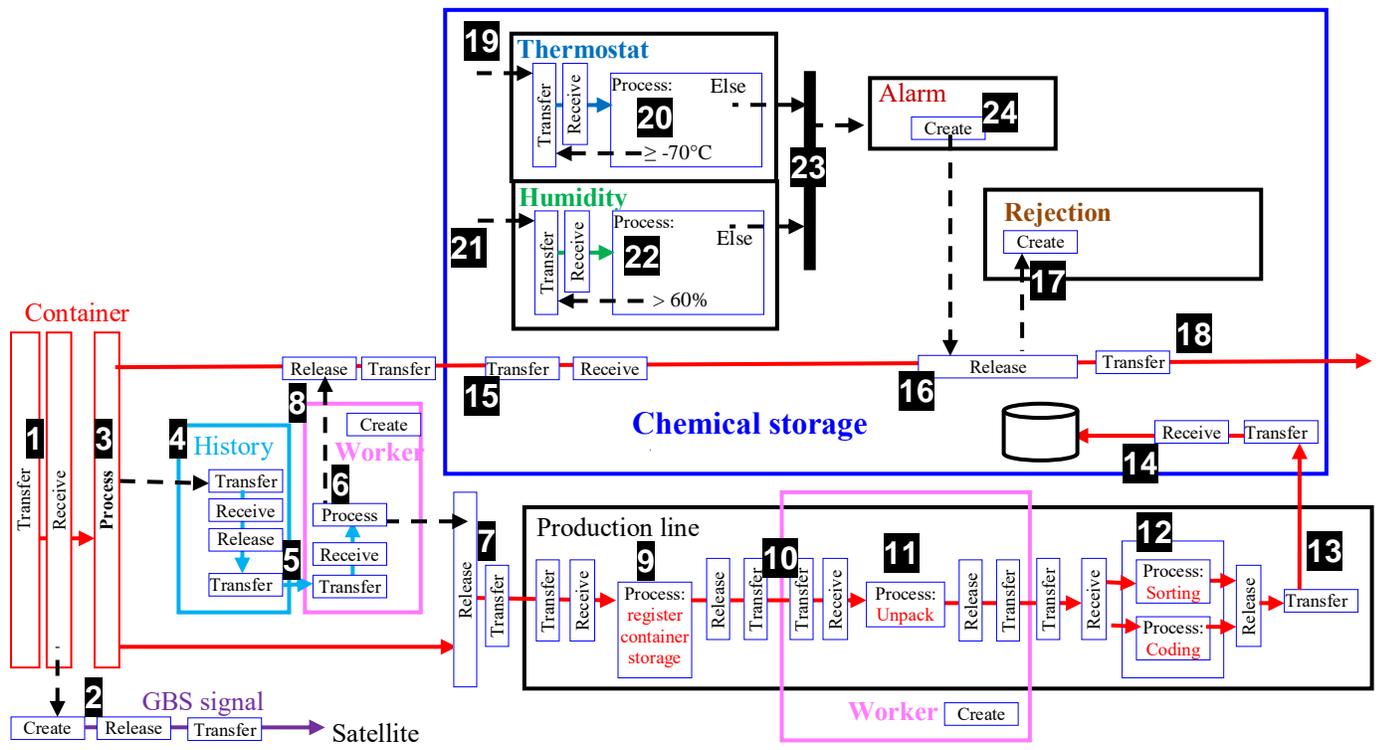

Fig. 6 The static TN model of the smart factory

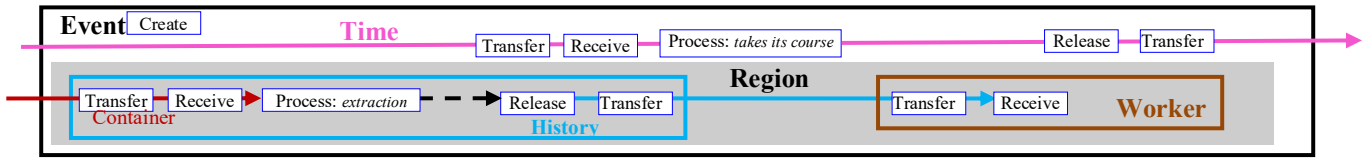

Fig. 7 The event *The container history is extracted and sent to the supervising worker*

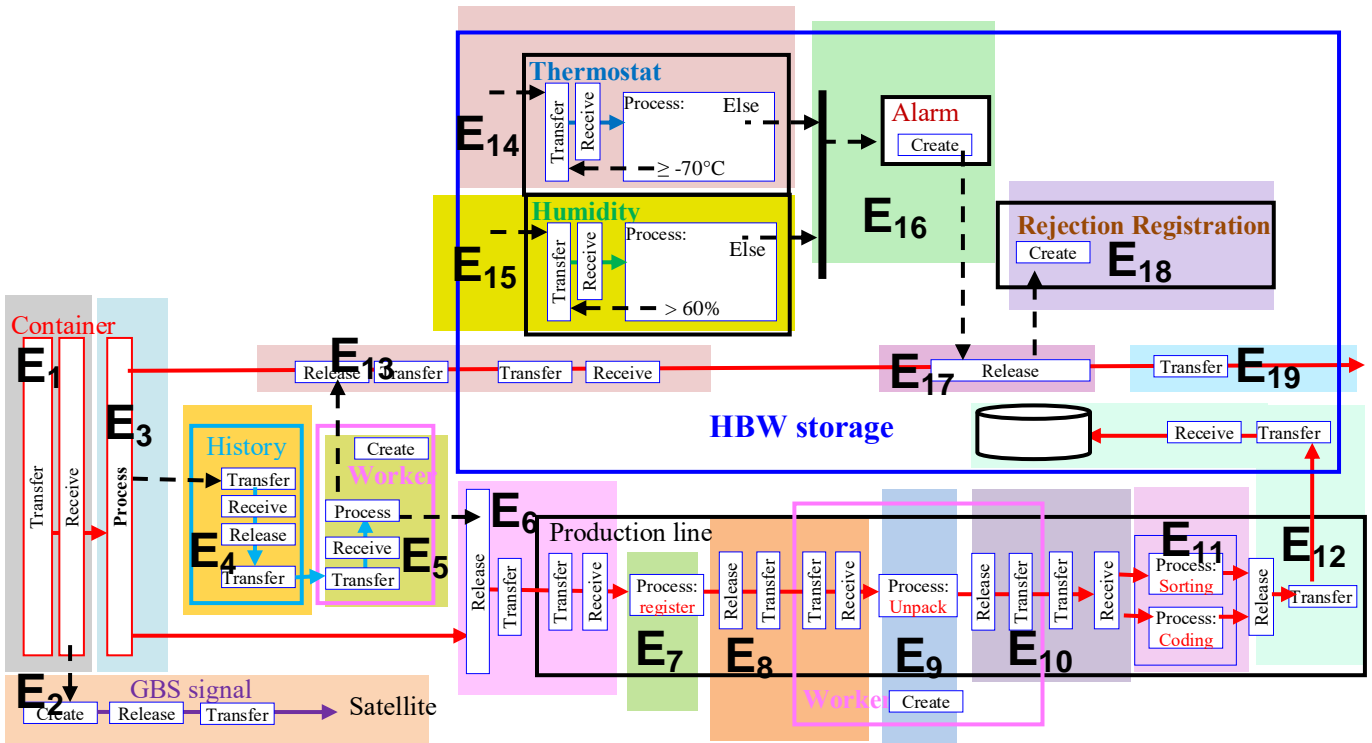

Fig. 8 The dynamic model of the smart factory

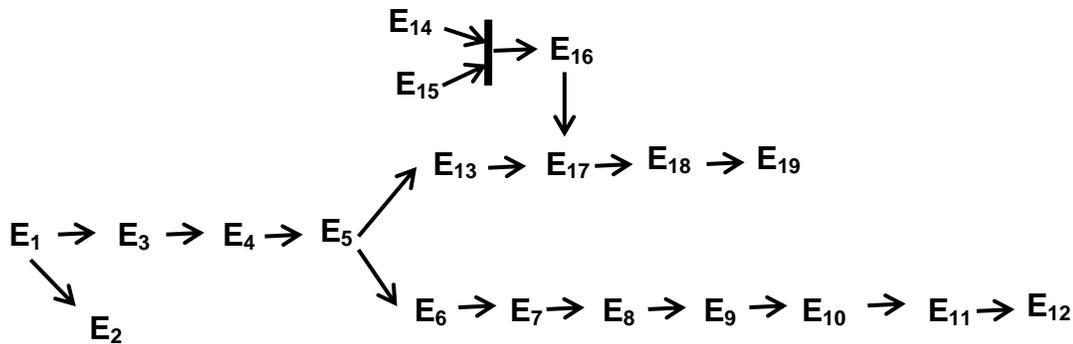

**Fig. 9 The chronology of events in of the smart factory**

In philosophy, the so-called *noumenon* is said to refer to the *thing-in-itself* as opposed to what is called the *phenomenon* (i.e. the thing as it appears to an observer; Kant). Noumenon is not directly accessible to our senses as the true nature of the independent thing in reality. According to Kant, man can only speculate and can never directly connect with the noumenon. Kant calls the real world without time and space the noumenal realm.

In TM, there are *two levels* of reality of a thing:
- **Type-a** thing in the existing level that is ***directly*** accessible to our senses (e.g. a man).
- **Type-b** thing in the existing level that is *indirectly* accessible to our senses (e.g. traffic).

In both cases, when a thing does not exist, it is in the subsisting level as potentiality.

Another example of types of things in philosophy is the long scholarly dispute about the existence of universals—often conceived, in opposition to particulars, as entities. Whitehead's process philosophy views universals as real potentials essential to the quantum processes modern science recognizes as fundamental in the nature of things [13]. Alexius Meinong (1853 – 1920) [14] held that things can be divided into three categories:
- *Existence*, or actual reality, which denotes the material and temporal being of an object.
- *Subsistence*, which denotes the being in a nontemporal sense (e.g. numbers).
- Absistence or being-given, which denotes being a thing but not having being [14].

Russell questioned the validity that such a thing as a *unicorn* denotes a reality and considered it as an "incomplete symbol" rather than a name of a thing.

The TM model agrees with both Meinong and Russell: *Subsistence* as a mode of reality includes these things that are mappable to existence. Because the *unicorn* has no possible independent od dependent existence, it is not an existing or subsisting thing.

We have already claimed that an ontological base for the TM model is the Stoic criteria of reality: a corporeal state for existence and an immaterial subsistence. Some philosophers mocked the Stoics' distinction between existence and subsistence as being an "absurd" notion [11]. However, the *two levels* of existence and subsistence provide a new ontological outlook for the Stoic ontology.

In TM, both existence and subsistence are modes of things in reality. The Stoic introduced subsistence in reality *abruptly,* and they meant existing incorporeal things. TM extends subsistence and identified incorporeal existing things in terms of emergence from the subsistence level. In particular, the type-b *subsisting* thing has a mode of reality because it can exist through being dependent on an already existing thing; for example, traffic is real through cars and roads. These issues will be illustrated in the next section where we will try to strengthen the level of reality-ness of type-b things of subsistence.

## IV. ONTOLOGY OF TM: EXAMPLES

The aim of this section is to provide a coherent picture of *what is reality* because TM modeling is supposed to represent a "portion of such reality." This attempt to clarify the blurred frontiers of the TM model is not purely metaphysical; rather, it is a necessary supplement to complete the modeling process by understanding this "portion of reality" in the broadest possible sense of the term. The effort is not a quest for *truth* about reality but an endeavor to add a level of conceptual intelligibility that strengthens the rationalization of the TM modeling approach.

TM CM brings two modes of reality of things: staticity (lack of change) and dynamism. The existence of a thing implies manifestations of dynamism in its previously subsisting mode of reality. Type-b subsisting things are dependent things that are actualized through being "carried on" by **already existing** things, and these things are incorporeals that depend on a body without themselves being bodies (Stoic terms). They do not have independent existence. We ignore other types of things (e.g. static-level things that cannot be actualized; e.g. square circles). Hence, herewith, the static level will refer to potential things only: type-a and type-b.

In the following two sections, to provide clarity to the notions of subsistence, existence, and their things, we intentionally repeated illustrations as a way to highlight their different facets and manifestations.

*Examples*
**1)** As discussed in the traffic example in the introduction, in Fig. 2, if cars and roads exist, traffic comes into existence, even though it may be in subsistence for a while. Note that cars and roads are also (subsisting) "wholes" that come into existence from their constituents (all are material things). Such a process continues downward until reaching basic things that come into existence without emerging from constituents (e.g. big bang).

Consider *traffic* as an existing process concealed *in* physical cars, roads, and movements (the upper part of Fig. 2 of the introduction). When all cars stop (e.g. COVID-19 pandemic), the traffic subsists as an imprint of the process. In this situation, the traffic has not perished (dissolved); rather, it simply revered to its potential mode of reality, which is in subsistence.

**2)** As illustrated in Fig. 10, the immaterial traffic can be "perceived" in mind; otherwise, the dog in the figure would not hesitate to cross the road. Subsistence can be used as a scheme of composition of physical things (events). The traffic is a *real* thing despite being incorporeal or not being bodily (Stoics used these terms) because it emerges from material, existing things.

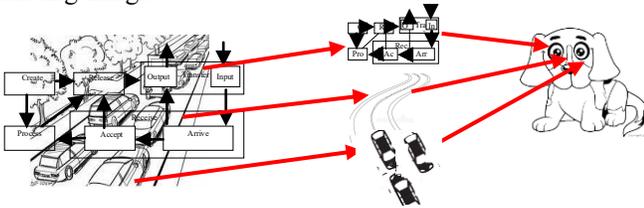

**Fig. 10 The three components of traffic are perceived. Traffic is represented by the static TM model**

**3)** In TM, subsisting things can be created, processed, released, transferred, and received. A *flock of birds* is a TM existing thing (the Stoic called it subsisting thing) that may be created (exists) and that subsists and reappears. As long as the birds are flying, the flock may *subsist* (the birds parted) to emerge into *existence* at any moment while in air. The birds and flying process already exist ready for the emergent phenomenon of flocking. The flock acts and moves as one group "like a drunken fingerprint across the sky" (American poet Richard Wilbur). A fish assemblage even has a leader when it turns away from predators.

**4)** A relation between an existing thing and its subsisting mode is a running computer algorithm and its source that is "carried on" the symbols of its code (conveyers). In around 825 AD, Al-Khwarizmi created his algorithm of solving quadratic formulas. The algorithm *subsists* inside its Arabic text as an immaterial thing. Now, the algorithm can be "carried" into a code of a programming language.

The algorithm can be brought into existence by executing its binary code on a computer. The symbols of the code (Arabic, programming language, or machine language) are not the algorithm; rather, the algorithm is a subsisting thing that comes into existence upon running the program.

We can see in these examples that subsistence is *a state of reality that has not yet emerged into a total existence*. When a subsisting thing emerges into existence, it is real regardless of whether it is material or immaterial. The difference between subsistence and its existence is the *emergence* to occur. In the TM static level, we have a *suspending* reality that consists of things that are "ripe" for emergence and future existing. *Ripe* here refers to the default value of happening (e.g. suspended traffic). Hence, subsistence is an important state of reality to "tone up" itself. It is analogous to the "Laplace's demon" in a calculating state before picking out a new state of the universe.

## V. Ontology of TM: Elaboration

Basing type-b things (immaterial, *existing* things) on material, existing things justifies the claim that type-b things have a facet of reality. Nevertheless, such a claim is presented in "the mode of plausibility [that] can proceed on a tentative basis and need not present their assertions as categorical claims to truth" (Rescher, see [15]). Existence is a matter of *clothing* subsistence (potentiality), and the origin of a TM region's subsistence is the presence of the region within the (existing) event. It is said that humans are "clothed in flesh and woven of bones and nerves" [16]. In such conceptualization, the flesh, bones, and nerves are existential aspects, and the region is the preexistent structure (represented by the TM subdiagram) of a human. This preexistent human is *subsisting* as a fertilization recipe, and to emerge in existence, male and female gametes fuse, producing a diploid zygote. Such a prescription (represented as a TM subdiagram) is "carried on" males and females in reality just as traffic is "carried on" cars and roads. The prescription is not a mental fiction or an abstract notion but a (real) root for existing humans. With respect to the foundation of reality, we may debate the significance of the region versus the event (e.g. Coca Cola formula vs. Coca Cola bottles).

The creation of an event involves a region that generalizes the notion of form (Plato), and the region conceals itself within the event.

### A. Dual Modes of Reality

The thimac (and subthimacs) has dual modes of reality that work together: static (timeless) reality (reflects the "meanings" of things; e.g. what is it?) and dynamic event-based reality (reflects the existence of a thing). The origin of such dyads of reality is the classical philosophical view (e.g. Avicenna) that conceptualizes a contingent as existence (it is) and *essence* (what it is, definition). The German philosopher Paul Hartmann viewed reality (TM existence) as made up of a chain of temporal events, and, according to him, the existence of a leaf originated in the essence (TM region) of the tree, whereas the existence of the tree belongs to the essence (TM region) of the forest. Note that it is possible that a leaf is existing while its tree is subsisting.

In TM, the essence of an event is its region. For example, the fire event has the sensed site of fire and the effect of burning that cannot be sensed only by **previous experience,** as illustrated in Fig. 11.

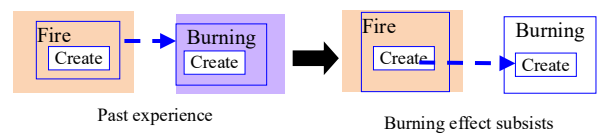

**Fig. 11 Burning effect is *known* from a *first occurrence*; thus, afterward, we don't have to experience the reality of burning to know it**

In the prehistoric *Croods* movie, the Croods family, first, knew the fire as "sun material," and after experiencing its effect, they knew what is burning. After that, the existence of fire means the subsistence of burning. The subsisting burning emerges in existence as soon as something (e.g. foot) touches the fire.

Ontologically, the ***first appearance*** of water in the universe is as *existent* thing that causes the preservation of its static region (quiddity or essence). Afterward, the region is infused into existence. The region becomes the blueprint, formula, or recipe for any further production of water. Inside this recipe, the thing of that region *subsists* just as an algorithm exists in a program.

### B. First Occurrence of Region

The static level reflects (mirror image) the structure/configuration of reality and echoes the whereabouts of events (regions). In TM, the static-level region possesses the [Plato's] form and attributes [17] and, additionally, the *actions* of the event represented statically. This mental-independent description originated from the *first occurrence* of the event (first temporal event), analogous to "dreams of the night are shadows of the day" [17]. The first occurrence of an event leaves an impression in a form that is replicated by the TM region, and this region remains after the first event itself has ceased to be, to be used to ignite (emerge) the same type of event in the future. Other occurrences of this same first event require no first region because "the region of the first event" has been registered in the static level and emerges with the new occurrences (see Fig. 12).

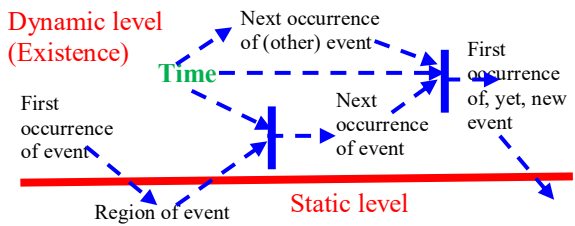

**Fig. 12 Regions in the static level is originated in existence**

According, the history (increasing past) of the universe is a series of first-time events that are repeated to create new first-occurrence events resulting from (encapsulated) regions and time. Regions persist when their events no longer exist. The human memory may be conceptualized as photographs of the regions of the past.

### C. Hierarchy: Cars and Roads Exist, Causing Traffic to Exist

The general thesis is the introduction a hierarchy (e.g. the traffic/cars/roads triangle) involves the emergence of existing things (events) from subsisting things (regions), which, in turn, emerge from the first occurrences of existing things (events). This hierarchy extends downward to basic things in a similar fashion to such basic ontology of quantum mechanics (*subatomic* particles or *regions* or points of *spacetime*) or chemistry (atoms and the bonds that occur between them). In a TM, basic things come into existence without emerging from constituents.

The *potentiality* of a thing is a wider notion than subsistence as a theoretical concept where a thing may **not** necessarily be based on existing things. Space debris was a potentiality even before the existence of space satellites. It became a subsisting thing when satellites launched and then existed with the appearance of dead satellites.

### D. TM Two Levels Overlay Each Other

The given TM examples give the impression that static and dynamic levels are separate. However, this picture emphasizes the characteristics of each level. The TM model reflects projected levels superimposed over each other and events serving as "ghosts" for regions. Dynamism appears as continuous successions of regions (as in movies).

Hence, existence and subsistence are like a double image impression (e.g. Rubin's vase), which is possible with a Gestaltic figure-ground perception. When viewing an event, spectators simultaneously perceive its region. The region has real subsistence, but such type of reality is "absently present" [17]. The mind can conceive quasi-real subsisting things purely in itself without considering their "existence," which is different from nonexistent (again, Rubin's vase).

### F. Subsistence Recurs in Philosophy

Subsistence is an important notion in the context of TM modeling because the *region* has a *subsisting* mode of reality and the *event* is an inhabitant of (physical) *existence*, the other mode of reality. In a TM, subsistence is the inscriptive world as the universe evolves, registering the creation of new things. It emerges as a potentiality from the addition of new things to actuality. Subsisting things are distinct from both actual things and mental things. Such an idea has frequently arisen in philosophy. For example, Gottlob Frege (1848—1925) claimed that declarative sentences are neither external *concrete* things nor mental entities of any sort. In this tendency, reference [18] introduced the "third world" of abstract, objective entities.

### G. Existence vs. Subsistence

Existence is real because of its physical and configuration form "presence," and subsistence is real because it is an element (along with time) in the composition of existence. Whenever there is an existing thing, there must also be a thimac that serves as its region in events. Subsisting things are immaterial, yet they are still imprinting (footprints) of existing things in reality.

Existing things (particulars or events) are perceptible ever changing things, whereas subsisting things are unchanging things and are derivative of existing things (e.g. structure, configuration source of an existing thing). Thus, existing things have 'more realness' than subsisting things. On the other hand, one can see existing things (actuality) only when they have subsisting region (potentiality). The origin of such a doctrine can be traced to Aristotle's matter and form, which are distinguished with generality (particular vs. universal) and modality (potential vs. actuality). However, the TM region is different from the classical notion of *form* in the sense that regions are not independent realities but are extractions from events (changes; philosophy in [19]: The real consists of a static everlasting *pre-existing frame*).

We speculate that existence and subsistence represent the two scales of macroscopic (perceived) and microscopic (not perceived) reality. The macroscopic scale of reality is the world of existence populated by events. The microscopic level of reality is the world of subsistence by regions (nets of thimacs).

*H. Events*

TMs treat activities and the so-called objects in a uniform way: as events. According to [19], an object is actually only a *relatively* stable entity and what one perceives (in a snapshot) is something like static snapshots from an underlying, ever changing process. In TMs, the "object" (event) is *partially stable* because the *create* (manifestation of existence) of its region persists in all slices (an "extension"), while other parts of its region (process, release, transfer and receive) are active causing slice-series with extended creation (e.g. the *same* "actor" goes through different regions). Thus, *creation* is extended (repeated in time) but process, release, transfer and receive are changing.

Objects are nothing more than long events (i.e. dynamically steadied processes, (Alfred North Whitehead (1861–1947) philosophy). In a TM, a process is an event or an assemblage of events, and existence is the flow of events. *Subsistence* is the reservoir of all the potential events in the *actual* domain (the thimac in which things and processes happen). Each event is built from a time thimac and subsisting *region* (subdiagram of the static description) with a logical (conceptual [20]) order. The TM static level is a strictly logical *Being* of thimacs. Regions emerge as "sparks of light" [17] of events (may be ongoing new formation of electrons and photon connections). Events are individuals of region types. That is, the events can self-repeat over the same region.

*I. Specification of Space*

The order/organization imposed by *potential* flows and the triggering mechanism are captured in a TM. The successive relations of earlier and later static actions are obeyed at the existence level. That is, potentiality is a restriction on actuality. The static world has some succession and order, yet it is timeless.

There is no "location" for the notion of *space* in the very thick jungle of thimacs. The place of a thimac is itself. If there is a "not Being," then it is a thimac without *create*. Einstein pictured space as an abstraction from relations among objects. In a TM, nettings of thimacs include things, actions, and flows as a thick "jungle" that replaces the notion of space.

*J. Negative Events*

TM modeling involves a vertically dynamic depiction over a *timeless* (static) image (of subsisting thing/process). In this sense, existence and subsistence are divergent of each other, where a subsisting thing is the not-yet "emergen-ized into existence" thing. This "not-yet" existence leads to subsistence and refers to a thimac being in subsistence and the use of Lupascian logic to represent a type of negativity [21]. The *existing* thing (event) is no longer in the exist stage, but it has taken residence in subsistence: "there is no absolute void in nature […] the absence of one thing is only possible by the more or less explicit representation of the presence of some other thing" [22]. The original existing thing's precise outlines and structure is now elsewhere: subsistence. One then remembers the thing and perhaps *expect* it to occur (encounter it) again. To think of the thing as "nonexistent is first to think the object [itself] and consequently to think it existent; it is then to think that *another reality*, with which it is incompatible, supplants it. *Only, it is useless to represent this latter reality explicitly; we are not concerned with what it is; it is enough for us to know that it drives out the object*" [22] (emphasis added).

Subsisting is also the negation of event (ing), for example, a not existing traffic process means it has reverted to the subsisting level even though cars and roads still exist. *Negative* (absence of) existence refers to a thimac being in subsistence. "The cat is not on the mat" indicates the *transfer and receive* of the cat to the mat is in the subsisting level (as illustrated in Fig. 13). Note that Bertrand Russell (1872 – 1970) solution for this negativity is the statement "There is a state of affairs incompatible with the cat being on the mat."

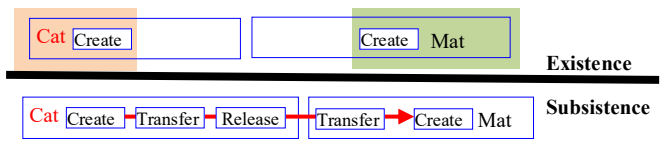

**Fig. 13 Negativity as represented in *the cat is not on the mat*. The *cat* and *mat* exist, but the move of the cat to be on the mat is not in actuality**

*K. Regions*

The region is the "seat" that provides a "position" for an event in the "existence sphere." The region is snapshots from an underlying, ever changing reality [19]. A region may combine with time to form a *dynamic* event. Subsistence is static (absolutely immobile), it emerges with *first-time* events. Note that existence implies both actuality and potentiality because the region is concealed in its event. Such a notion originated with Aristotle: that we cannot identify potentiality without reference to the corresponding actuality. Distinct regions represent distinct events; however, distinct events may have the same region.

Suppose that *amino acid* came into existence *for the first time* in nature by lightning strikes. The subsistence thesis claims that such a first event additionally produces an immaterial inscription (form, region) of the structure of amino acid as a *potential* thing (subsisting) in the inventory catalogue of reality. Thereafter, a subsisting region of amino acid precedes the event of creating amino acid; that is, the next time of occurrence (existence) of amino acid is a composition of such region and time. Furthermore, we advocate that the static TM model represents such a subsistence of things.

If an event happens for the first time, then its region (micro-scratches, a Bergsonian term) is "registered" as a potentiality. Suppose that the universe comes into existence by an event that contained a single, hot, dense point, the so-called, *big bang*. The big bang event created (come into *existence*) hydrogen, helium, and lithium to form heavier elements *for the first time*. This is accompanied by

"registering" (creating *subsisting* regions) of hydrogen, helium, and lithium as elements of the universe. Accordingly, the static inventory catalogue of reality is supplemented with regions of these events. Afterward, creating hydrogen, helium, and lithium, time after time, does not come into existence directly from the big bang; rather, they are created by time "activating" their regions. This discussion implies that potentiality (subsistence) "emerges" from actuality when the latter happens for the first time. The subsistence reality is only in relation to existence, which alone detected reality.

Such a claim is based on logical possibility and noncontradictory and empirical evidence in world history that not all things came into existence directly from the big bang. The universe has since increased its list of things. However, if we accept the big bang event, then nothing exists *in and of* itself except the big bang. Everything exists only inasmuch as it links to already existing things. Even in religious beliefs, only Adam and Eve were created anew, then humans descended from processing DNA sequences as code for the book of life, decipherable in terms of information "carried on" protein semantic units.

The repeated re-existence of things dictates preserving their formulas (imprints) to build them from other things. This "file" of preservation *ought* to be in some mode of the subsistence level. Because all subsisting things are in the static world and all existing things are the regions' counterparts on the dynamic level, regions *exist* (not subsist) only in the interior of events.

### *L. Speculation About TM Region-based Quantization*

We can consider the nature of regions of events and the static property using quantum theory in [23]. *Quantization* is the property of appearing only in lumps of discrete sizes. There is an ultimate lump or "atom"; that is, a *photon* that is a limit to how thinly a thing (e.g. light, gold) can be evenly spread, beyond which they cannot be subdivided without ceasing to be a thing. In this case, "for quantities like distance (between two atoms, say, the notion of a continuous range of possible values turns out to be an idealization. There are no measurable continuous quantities in physics […] If everything is quantized, how does any quantity change from one value to another? How does any object get from one *place* to another if there is not a continuous range of intermediate places for it to be on the way?" [23].

Is this state the level of stacity and subsistence as in TM modeling? The region "is there" as a subsisting thing with static process (change) and static move (static release, transfer and receive). Only the structure of the region of things and actions can be detected without the continuing feature of existence.

Reference [23] then discussed the quantum slit experiments that demonstrate that photons when projected through slits, display interference patterns on a screen even if they come through the slits separately, one by one. How is it possible for a single photon to "interfere" with itself? Or,

"When a single photon at a time is passing through the apparatus, what can be coming through the other slits to interfere with it?" [23] It appears that photons come in two sorts: tangible photons and shadow photons. Tangible photons are detectable with instruments, whereas shadow photons are intangible (indivisible)—detectable only indirectly through their interference effects on the tangible photons. Reference [23] discusses the issue of things and their shadows.

## VI. INFORMATION

The concept of information has been associated with many concepts, including knowledge, messages, symbols, and signals, among others. For example, in [24], three meanings of "information" are distinguished: "as-process," "as-knowledge," and "as-thing." Traditionally, information is often conceptualized as part of a hierarchy of data-information-knowledge-wisdom, which bases information on data. However, "it is not an exaggeration to say that there is no consensus on what information really is" [25].

In this paper, the view that information is an element of reality is of special interest (see [26] for many references). Information can be measured, quantified, transformed, observed, and used. It is a property of matter and the physical world, and it is the third constituent element of reality [27]–[28].

In this paper, we avoid physics-based conjectures (e.g. information mass, dark matter, elementary particles, quantifiable information) and concentrate on conceptual modeling in developing information ontology based on the previous sections. In a TM, information is created as footprints of events. Events provide "content" to information. The same TM region leads to different events and, hence, different information.

### *A. TM Information*

The example in Fig. 14 illustrates the TM conception of information in terms of the event *The bulb being turned on* at a certain time. In a TM, *information* refers to information *about* an event.

In Fig. 14, **after** the event *The bulb being turned on (now)* occurred, the event footprint is "registered" (downward green arrow) as subsisting information (pink 1) at the static level. We ignore here what happened before this event because this is understood from previous sections (i.e. region and time emerge as an event). Note that such information *is about* the event (2) (after the event occurrence) with its region (3) and time (4). Thus, the event creates a reflection of itself that is registered as a subsisting region. This "event of event" is a type-b (immaterial) event.

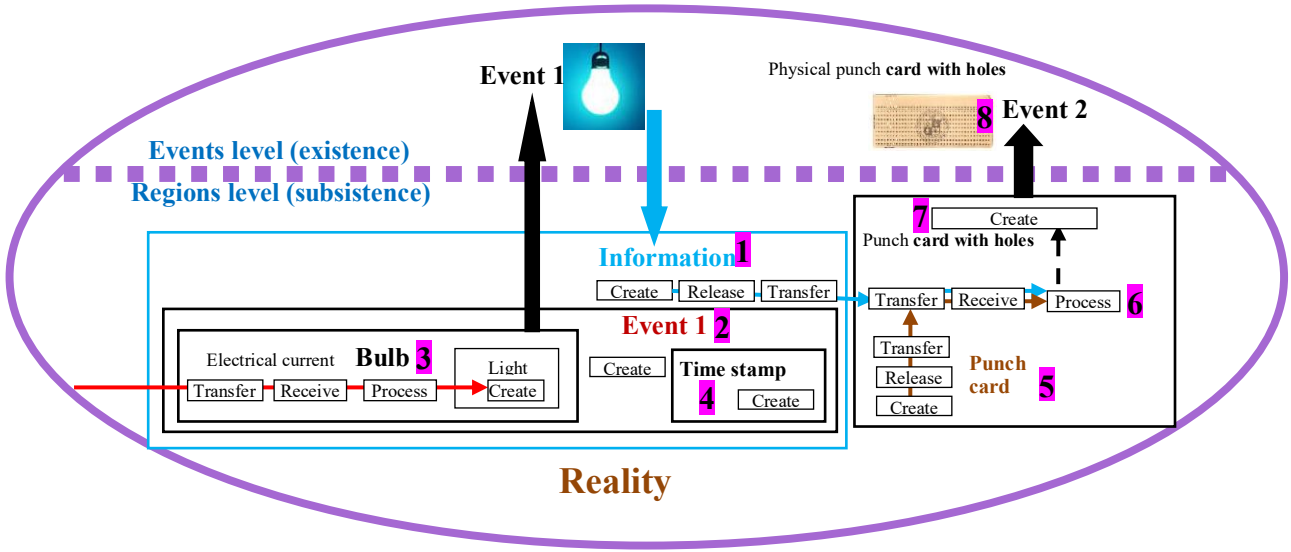

**Fig. 14** The information about the event *Turning a bulb ON* is preserved in subsistence to be "carried on" a punch card with holes.

This static region) of the information cannot come into existence by itself to be "carried on" by a punch cared (5) processed (6) along with the information to create a punch card with holes (7). The punch card with holes can now exist as event 2. The point is that information has dependent existence, just as the traffic discussed previously.

Information is something about events; that is, it is about existing things. Its region has subsisting nature that can be brought into existence only through being "carried on" (to existence) by other things. According to such an approach, a bit is not information; rather, it is a "carrier" of information, just as a punch card.

### B. Nature of Information

Consider the famous Aristotelian signet ring pressing its form into wax. This pressing is an *event*. Our thesis is that events self-inscribe themselves in reality. The inscription (information) has dependent existence ("carried on") in reality. In other words, the world of happenings *mirrors* itself through a twin world of inscription. This inscription world is formed from footsteps and tracks of events. Events are 'projected' on the static thimacs dimension. Such a notion of inscriptional things is known in philosophy. For example, Gilles Deleuze (1925–1995) stated that the human face (the carrier), has inscriptional unstable aspects that mark feelings and emotions.

The form in the wax is a footprint of that event (ring-pressing) carried by the wax. At the quantum level, we "get" only footprints or tracks of the actual photon. Events can also leave immaterial footprints in its surroundings. Unlike the ring and the photon footprints, events seldom leave perceived tracks in the environment. Such event tracks (information) are real and mostly static and immaterial. Events are existing things, and event footprints are their tracks. In TM modeling, we *capture* event footprints (information): the ontological structure of the existing event in diagrammatic form as illustrated in Fig. 15.

So a (static) region becomes an event when it entangles with time. Information is created in a reverse process: an event is "frozen" as a region in the static level. Such information cannot return to the existence level independently, only by "riding" on an actualized noninformation thimac. In general, an event reflects the region that is invariant under the TM sequences of actions that cause changes.

Fig. 16 illustrates the situation when the event *It is cloudy (now)* triggers the creation of its information captured in the subsistence level. The figure shows the propagation of *It is cloudy (now)* from the event to the observer's brain. First, *It is cloudy* is carried on some rays to trigger event 2, actual rays mounted by the information. The information carried on the *real* physical rays reaches the eye to be extracted from the rays and processed to create electrical signals loaded with information that travel to the brain. Thus, the eye is a "change station" where information is carried on by electrical signals instead of physical rays. In the brain, the electrical signals are processed to extract the information. The extracted information then has some type of immaterial "brainy existence" involved in the immaterial process of thinking. Immateriality is the negativity of materiality. It means nonexistence or, in a TM sense, subsistence.

This example reflects the fundamental problem of communication in reproducing at the destination, either exactly or approximately, the information originating at a source. Information is carried on along the flow but never exists by itself.

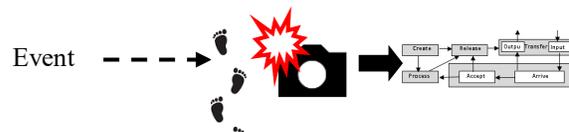

**Fig. 15 TM model as a seizure of the immaterial footprints of events**

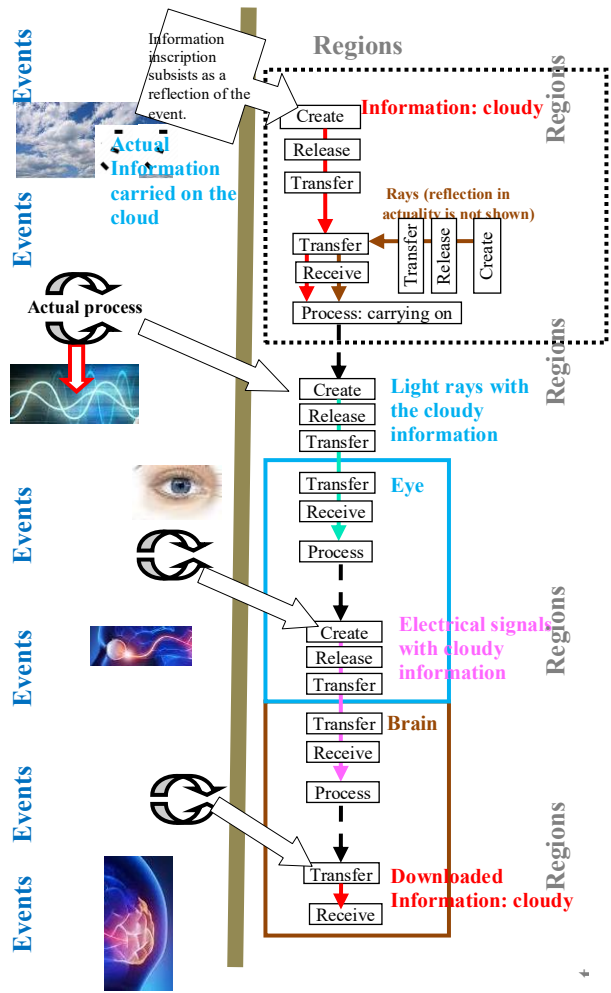

**Fig. 16 Information is "carried on" until loaded in the brain.**

In a TM, "information" corresponds or refers to the original event's specification (region and time). Information reflects events (existence) approximately as (ignoring stability issue) a mirror reflects events. The difference is that the mirror "replicates" the event, whereas the region in the TM static level "encodes" the event in a form of subsisting nature.

Information is a "subsisting event" and as "real" as the mirror's images in practical purposes on an everyday basis (e.g. car's mirror). In Fig. 16, information has a nomadic character and dependency on its carrier to be transported through existence. It is able to act when it arrives at the brain (making decisions according to what happened in the original event).

In this description, the information mirrors the "real material" events. Information would be a replica of the event that occupies an alternative real world of subsistence. This seems to bring into consideration a similar semantic broadening of real being that has a certain nonphysical being such as mirror images. The information is different from the event, but it preserves its real being in a subsistence form that "moves" (by being carried) independent of the original event.

Reference [29] claims that "a mirror image is a *form* […] is an entity different from the material object [as] the indeterminate bodies […] the rainbow […] The mirror image is not a form in-forming the matter of the mirror—it exists in it as a mere species having immaterial being" (emphasis added). The information discussed in this section is not Shannon's bits which are mere carriers of information. It "informs" about the where and when actions occur. Information is a subsisting thing that becomes real when carried on by an existing thing (e.g. punch card).

## VII. CASE STUDY

This section presents the TM re-modeling of a financial service loan brokering company with an event-driven brokering system developed in [30]. According to [30], the conceptual model is based on event-driven architecture motivated by a lack of standardization in expressing the event-processing directives in event-driven systems.

The involved event-driven architecture provides broadcast events, publishing events as they occur and monitors relationships at each event. There is no explicit definition of an event, but in [31], an event is defined as "anything significant that happens or is contemplated as happening […]. It is contemplated as happening because it could be a fact becoming true or could be a transition of an entity in the real world." For example, a trade order has been issued, an aircraft on a specific flight has landed, sensor data has been read; or it might be monitoring information about IT infrastructure, middleware, applications, and business processes [30]–[31].

### A. Description of TM Static Model

Fig. 17 shows the static model of a financial service loan brokering company. Few processes were included (e.g. informing the lender about the borrower's negative response) to reduce the complexity of the diagram to one page. In Fig. 17, two shaded ovals represent the borrower and lender. The area outside these ovals represents the loan brokering company. Accordingly, the modeling process starts when the borrower creates and sends a request for a loan (points 1 and 2). When the company processes the borrower's request (3), the result leads to one of two directions to follow.

At point 4 in Fig. 17, the ID (6) of the request is extracted from the request. Note that the extracted ID starts with transfer/receive to indicate the appearance of the ID in the company's system from its concealed state as part of the request. This situation is analogous to the arrival of a taxi, which implies the arrival of the passenger. Then, the ID flows (6) to a procedure that processes (7) with a file that includes the of borrower's history (8). For simplicity sake, we do not include such a procedure in its own box. The result of this process is extracting the borrower's history record (9). This record is examined (10).

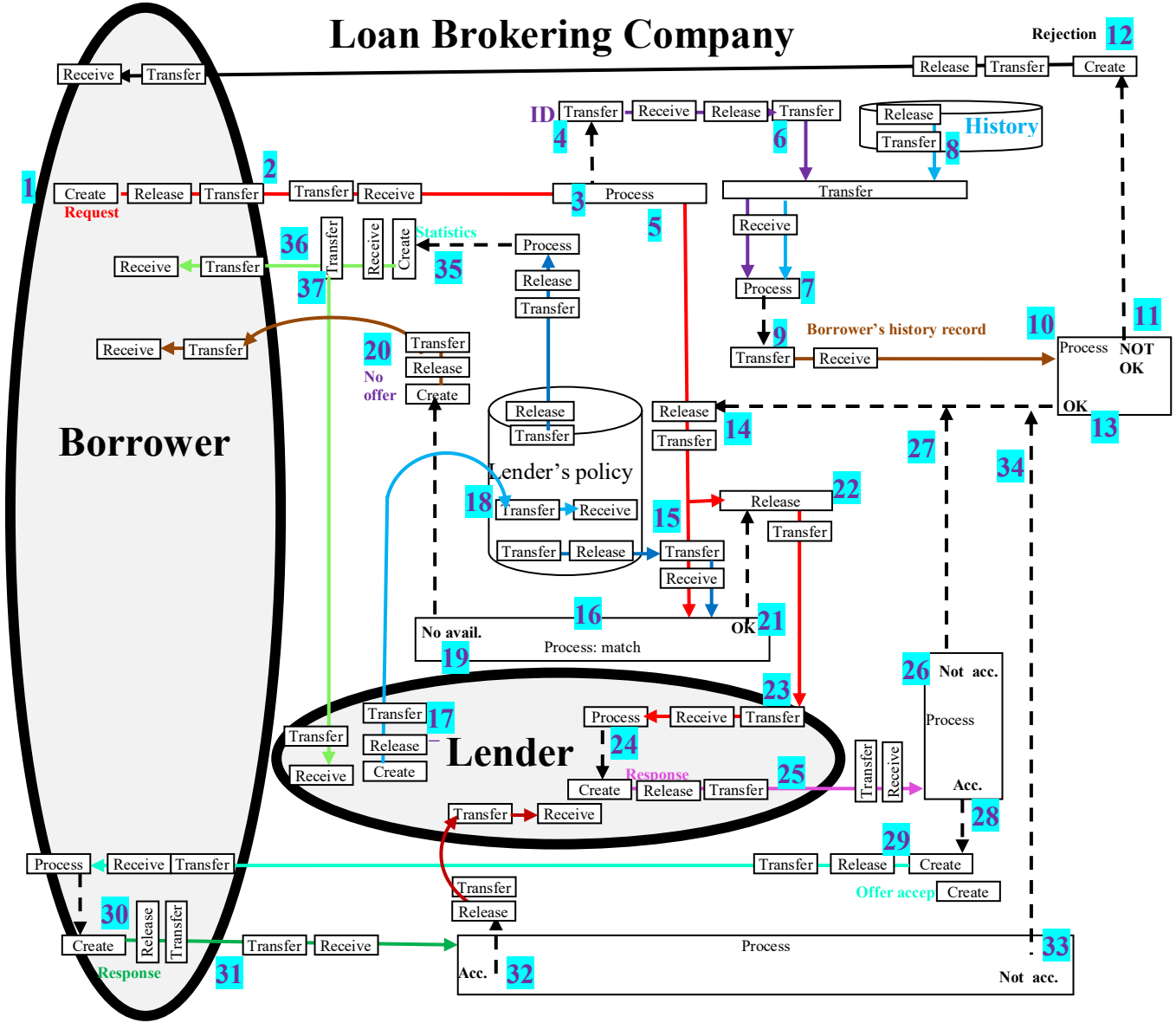

**Fig. 17 TM static model of the case study**

If the result is not OK, (11) then a rejection message is sent to the borrower (12). If the result is OK, the borrower's request is forwarded to the next state of processing (5 and 14).

The borrower's request flows (15) to be compared with records in the lender's policy (16), supplied by the lender (17 and 18). If the borrower's request details do not match any policy in the file (19), a "no offer" message is sent to the borrower (20). If a lender's policy is found (21), then the request is sent to that lender (22 and 23). The lender processes the request (24) and sends the response to the company (25). If the lender's response is negative (26), then the borrower's request is processed again to find another lender (27, 14, 15, and 16). If the lender's response is positive (28), then an "offer acceptance" message (29) is sent to the borrower.

The borrower processes the "offer acceptance" message and sends the corresponding response (30 and 31). In the company, if the borrower's response is positive, then this news is communicated to the lender (32); otherwise (33), the borrower's request is processed again to find another lender (24, 14, 15, and 16). Note that the lender's policy file can be processed (35) to create statistics that are sent to the borrower (36) and the lender (37).

*B. Dynamic Model*

Fig. 18 shows the dynamic model, and Fig. 19 displays the chronology of events. Note the role of event $E_{18}$ in providing information to the borrower. No further discussion for brevity.

14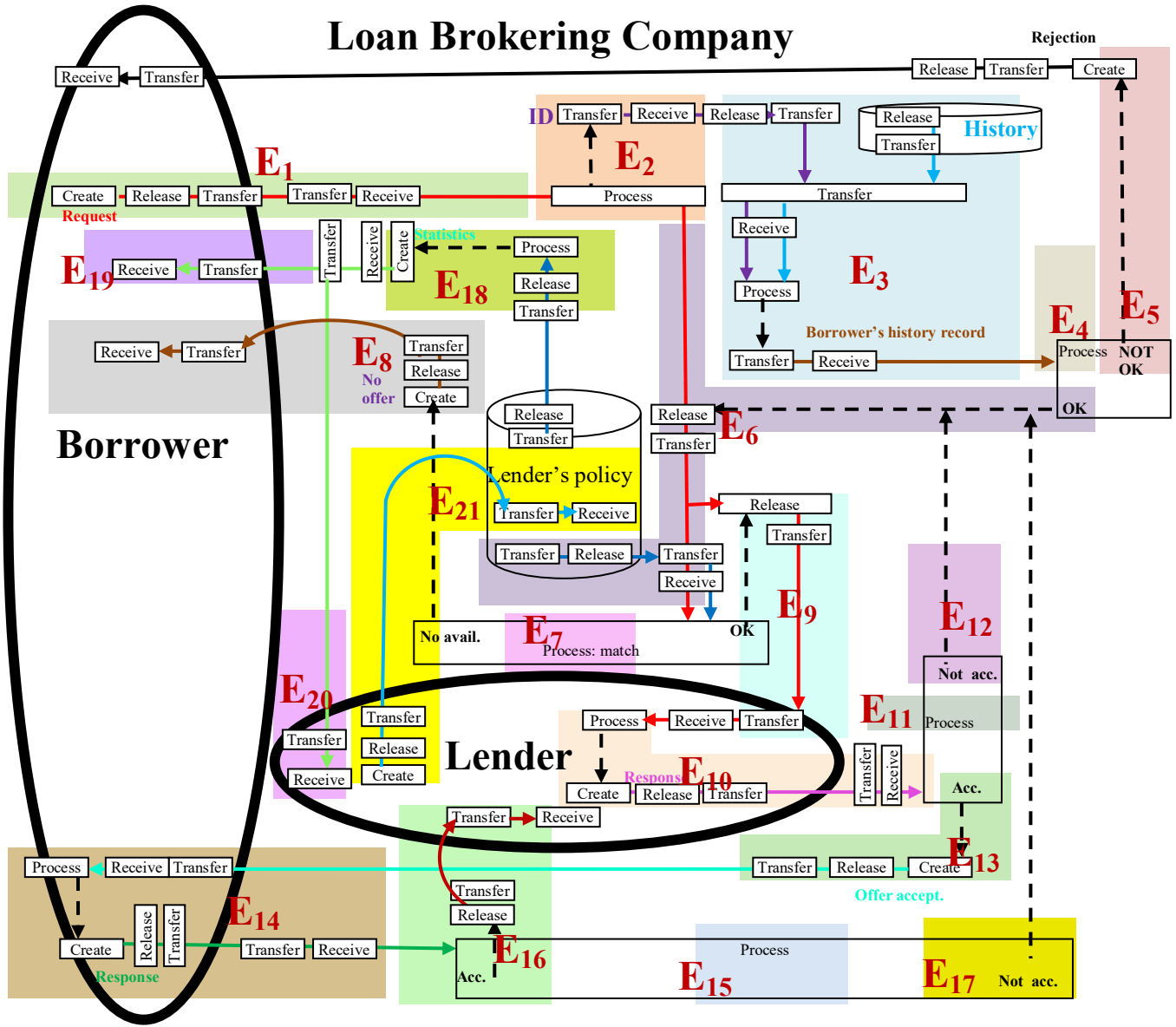

**Fig. 18 TM Dynamic (events) model of the case study**

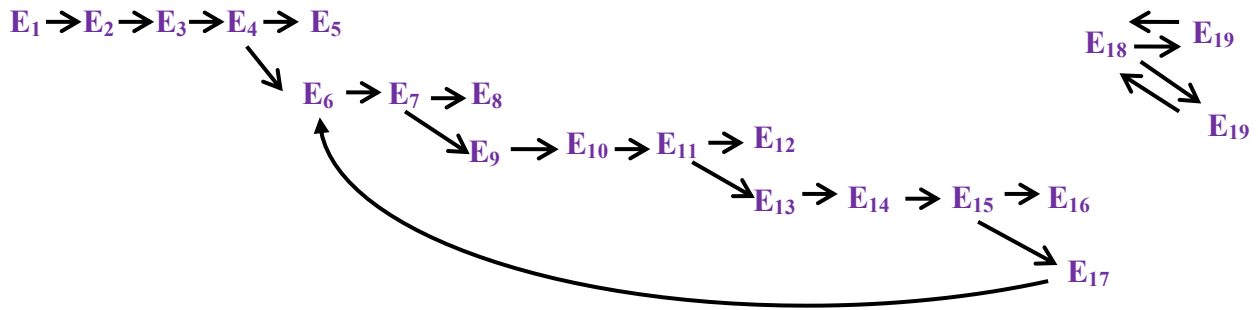

**Fig. 19 Chronology of events of the case study**

## VIII. Conclusion

This paper is an attempt to clarify the blurred frontiers of the TM model. It is not purely a metaphysical study; rather, it is a necessary supplement to complete the modeling process by understanding a "portion of reality" in the broadest possible sense of the term. The underlying thesis for such an effort is that conceptual modeling in software engineering has reached a level of maturity to further develop by defining basic concepts, such as the important notion of "reality" even though philosophy is still a foundation. The motivation for this is the difference between CM and philosophy in terms of objectives and method. For example, in CM, the aim of modeling efforts is not a quest for *truth* about reality but instead an endeavor to add a level of conceptual intelligibility that strengthens the rationalization of the modeling process.

Accordingly, the ontological explanation in this paper gives a reasonable explicit justification for claiming that the two-level TM has a mind-independent representation of reality. It also supports the Stoic claim of two-mode reality. The given TM ontology involves a two-level subsistence/existence scheme originating in the Stoic modes of reality. The categorical structure of TM modeling has two kinds of things and two modes of reality.

With respect to TM modeling, we can now argue that the static logical form of TM representations resembles the ontological structure of reality. Nevertheless, at this stage of research, future research needs to strengthen the proposed ontology.

## References

[1] T. Jonsson and H. Enquist, "Phenomenological ontology guided conceptual modeling for enterprise information systems," in *Advances in Conceptual Modeling*, ER 2018. C. Woo, J. Lu, Z. Li, T. Ling, G. Li, and M. Lee, Eds. Edinburgh, UK: Springer, Cham, 2018, vol. 11158.

[2] R. Lukyanenko and R. Weber, "A realist ontology of digital objects and digitalized systems," *Digit. First Era — a Joint AIS SIGSAND/SIGPrag Workshop*, pp. 1–5, 2022.

[3] R. Lukyanenko, V. C. Storey, and O. Pastor, "System: A core conceptual modeling construct for capturing complexity," *Data Knowl. Eng.*, vol. 141, p. 102062, Sep. 2022.

[4] R. Lukyanenko, J., Parsons, V.C., Storey, B.M., O. Samuel, Pastor (2023). Principles of Universal Conceptual Modeling. In: van der Aa, H., Bork, D., Proper, H.A., Schmidt, R. (eds) Enterprise, Business-Process and Information Systems Modeling. BPMDS EMMSAD 2023. Lecture Notes in Business Information Processing, 169–183, vol. 479. Springer, Cham.

[5] N. Guarino, G. Guizzardi, and J. Mylopoulos, "On the philosophical foundations of conceptual models," *Inf. Model. Knowl. Bases*, vol. 31, pp. 1–14, 2020.

[6] S. Cranefield and M. Purvis, "UML as an ontology modelling language," in *Proc. 16th IJCAI*, 1999.

[7] G. Guizzardi, G. Wagner, and H. Herre, "On the foundations of UML as an ontology representation language," in *Engineering Knowledge in the Age of the Semantic Web*, EKAW 2004. E. Motta, N. R. Shadbolt, A. Stutt, and N. Gibbins, Eds. Berlin, Germany: Springer, 2004, vol. 3257, pp. 47–62.

[8] R. Martinelli, "Realism, ontology, and the concept of reality," *Etica & Politica*, vol. 16, pp. 526–532, 2014.

[9] S. S. Al-Fedaghi, "In pursuit of unification of conceptual models: Sets as machines," preprint arXiv, 2306.13833, Jun. 2023.

[10] S. S. Al-Fedaghi and G. Aldamkhi, "Conceptual modeling of an IP phone communication system: A case study," *Int. J. Interdiscip. Telecommun. Netw.*, vol. 13, pp. 83–94, Jul. 2021.

[11] V. de Harven, "The Coherence of Stoic Ontology," Ph.D. thesis, UC Berkeley, California, May 2012.

[12] M. Winter and M. Reichert, "BPMNE4IoT: A framework for modeling, executing and monitoring IoT-driven processes," *Future Internet*, vol. 15, p. 90, 2023.

[13] L. Gibson (2008) Ontology of the shadow. [Online]. Available : http://www.lennygibson.com/wp-content/uploads/2012/11/Ontology-of-the-Shadow-full-paper.pdf

[14] H. Sack (2021) Reality according to Alexius Meinong. [Online]. Available: http://scihi.org/reality-alexius-meinong/ See also A. Meinong, "The theory of objects," translated by I. Levi, D. B. Terrel, and R. Chisholm, in *Realism and The Background of Phenomenology*, R. M. Chisholm, Ed. Illinois: The Free Press of Glencoe, 1960.

[15] A. Weekes, "After Whitehead, Rescher on process metaphysics," in *Die Deutsche Bibliothek, ontos verlag*, M. Weber, Ed. Heusenstamm, 2004.

[16] P. G. Boersma, "The context of Augustine's early theology of the Imago Dei," Ph.D. Thesis, Durham University, Durham, UK, 2013.

[17] M. Uršič, *Shadows of Being: Four Philosophical Essays*, Newcastle, UK: Cambridge Scholars Publishing, 2018.

[18] J. L. Falguera and C. Martínez-Vidal, "Abstract Objects," *Stanford Encyclopedia of Philosophy*, E. N. Zalta & U. Nodelman, Eds. Aug. 2021. [Online]. Available: https://plato.stanford.edu/entries/abstract-objects/

[19] F. Fischer, "Bergsonian Answers to Contemporary Persistence Questions," *Bergsoniana*, vol. 1, Jul. 2021.

[20] J. Marek, "Alexius Meinong," *Stanford Encyclopedia of Philosophy*, E. N. Zalta & U. Nodelman, Eds. Fall 2022. [Online]. Available: https://plato.stanford.edu/archives/fall2022/entries/meinong/

[21] S. Al-Fedaghi, "Lupascian non-negativity applied to conceptual modeling: Alternating static potentiality and dynamic actuality," preprint arXiv, 2210.15406, Oct. 2022.

[22] H. Bergson, "The cinematographical mechanism of thought and the mechanistic illusion: A glance at the history of systems, Real becoming and false evolutionism," in *Creative Evolution*, translated by A. Mitchell. New York: Henry Holt and Company, 1911, pp. 272–370.

[23] D. Deutsch, *The Fabric of Reality*, New York: Penguin Books, 1997.

[24] M. Buckland, "Information as thing," *J. Am. Soc. Inform. Sci.*, vol. 42, pp. 351–360, Jun. 1991.

[25] M. Burgin and R. Mikkilineni, "Is information physical and does it have mass?" *Information*, vol. 13, p. 540, 2022.

[26] R. Krzanowski, "Towards a Formal Ontology of Information Selected Ideas of Krzysztof Turek," *Zagadnienia Filozofoczne w Nauce*, vol. 61, pp. 23–53, 2016.

[27] M. Burgin and H. R. Krzanowski, "World Structuration and Ontological Information," *Proc. IS4SI Summit 2021*, 2022.

[28] R. Landauer, "The physical nature of information," *Phys. Lett.*, 1996.

[29] L. Lička, "What is in the mirror? The metaphysics of mirror images in Albert the Great and Peter Auriol," in *Senses and the History of Philosophy*, B. Glenney and J. F. Silva, Eds. London: Routledge, 2019, pp. 131–148.

[30] G. Sharo and O. Etzion, "Event-processing network model and implementation," *IBM Sys. J.*, vol. 47, 2008.

[31] M. Edwards, O. Etzion, M. Ibrahim, S. Iyer, H. Lalanne, M. Monze, M. Peters, Y. Rabinovich, G. Sharon, and K. Stewart, "A Conceptual Model for Event Processing Systems," *IBM Corp. Redbook*, 2010.